\RequirePackage{fix-cm}
\documentclass[smallcondensed]{svjour3}                     

\usepackage{graphicx,color}
\usepackage{moreverb}
\usepackage{amsmath}
\usepackage{amsbsy}
\usepackage{amsfonts}
\usepackage{amssymb}
\usepackage{booktabs}

\begin{document}

\title{Pore-scale modeling of fluid-particles interaction and emerging poromechanical effects}

%

\author{Emanuele Catalano         \and Bruno Chareyre         \and Eric Barth\'elemy}
\institute{Emanuele Catalano \at
	      Grenoble INP, UJF, CNRS UMR 5521, 3SR lab\\
	      BP53, 38041 Grenoble Cedex 9, France \\
	    \email{ema.catalano@gmail.com}  
	    \and
	   Bruno Chareyre \at
              Grenoble INP, UJF, CNRS UMR 5521, 3SR lab\\
	      BP53, 38041 Grenoble Cedex 9, France\\
              \email{bruno.chareyre@grenoble-inp.fr}           
           \and
	Eric Barth\'elemy \at
	      Grenoble INP, UJF, UMR CNRS 5519, LEGI\\	      
	      BP 53, 38041 Grenoble Cedex 9, France
}

\maketitle

\begin{abstract}
A micro-hydromechanical model for granular materials is presented. It combines the discrete element method (DEM) for the modeling of the solid phase and a pore-scale finite volume (PFV) formulation for the flow of an incompressible pore fluid. The coupling equations are derived and contrasted against the equations of conventional poroelasticity. An analogy is found between the DEM-PFV coupling and Biot's theory in the limit case of incompressible phases. The simulation of an oedometer test validates the coupling scheme and demonstrates the ability of the model to capture strong poromechanical effects. A detailed analysis of microscale strain and stress confirms the analogy with poroelasticity. An immersed deposition problem is finally simulated and shows the potential of the method to handle phase transitions.
\end{abstract}


\maketitle

\section{Introduction}

\label{Intro}
The description of the mechanics of saturated porous media is a problem of great interest in engineering. Various approaches have been attempted in order to model the behaviour of such systems, the complexity of their structure and the interaction between the solid and the fluid (liquid, gas) phases. At the microscopic (sub-pore) scale, the solid and fluid phases occupy different portions of the spatial domain and interact at their common interface. Thus, the microscopic fields which describe the properties of constituents may be considered as continua within a single phase, while exhibiting discontinuities at the interfaces between phases.
A key feature of saturated porous media is the two-way coupling between the deformation of the solid matrix and the fluid pressure. This coupling is the source of the so-called \textit{poromechanical effects}, which govern for instance the consolidation process. Karl von Terzaghi was the first author to describe these effects, in the framework of continuum mechanics \cite{terzaghi1923}. Later, Maurice Biot adopted Karl von Terzaghi's ideas and extended the one-dimensional consolidation theory to the three-dimensional case for linear elastic porous solids \cite{Biot1941}.

The main objective of the work presented in this paper is to capture the poromechanical effects in a discrete numerical model of a granular material. The model will be formulated for the special case of incompressible phases. In this situation Terzaghi's equations are consistent with Biot's theory of poroelasticity, and the poromechanical effects are maximized \cite{Detournay1993}. We will restrict the study to quasi-static behaviour.
The choice of following a discrete approach in modeling granular materials became more and more popular in the last decades. Analyzing the behaviour of granular materials at the particles scale enables a better insight into phenomena which are governed by particles behaviour, and gives access to kinematic and static variables which are extremely difficult to measure in experiments. Hence the links between the microscale and the macroscale can be investigated. A popular method for discrete modeling is the discrete element method (DEM). Initially, the DEM has been developed without considering the effect of fluids, and was therefore restricted to dry granular materials. In the recent years, however, many efforts have been devoted to couple the DEM with models of fluid flow within the medium. The various methods that have been developed differ in the modeling techniques adopted for the description of the fluid flow and are briefly reviewed below.

\textit{Continuum-based} models use continuum formulations and coarse-grid meshing for the fluid part of the problem \cite{Kafui2002,Shimizu2004,Zeghal2004,Chen2007}. The coarse mesh defines subdomains of the DEM model over which the porosity and velocity of the solid phase are averaged and introduced as field variables of the continuum formulation. In turns, the solution of the continuum problem gives fluid velocity and momentum exchange between the phases at each node of the grid. The latter is discretized again with appropriate rules to determine forces to be applied on the particles of the DEM model. The solution can be obtained with classical numerical methods such as finite differences (FD) \cite{Kafui2002} or finite volumes (FV) \cite{Shimizu2004}.
Continuum approaches generally lead to affordable CPU costs, since the number of degrees of freedom (DOFs) associated to the fluid can be much smaller than the number of solid particles. They need a series of phenomenological assumptions and rely on empirical relations such as Ergun's relation \cite{Ergun1952}. A drawback is that they may require a calibration procedure for each new type of microstructure. More importantly, they are inherently unable to describe accurately the effects of fluid at the particles scale. This is because the variables of the fluid problem are averaged at the scale of the coarse-grid cells, whose size is generally larger than the average size of solid particles.

Inversely, \textit{microscale} models are based on a very fine discretization of the void space. A Navier-Stokes problem where the boundary conditions are prescribed at the surface of each particle is then solved using conventional techniques (e.g. the finite element method (FEM) \cite{Glowinski2001}), or particle based methods such as the Lattice-Boltzmann method (LBM) \cite{Boutt2007,Mansouri2009,Han2012,Lomine2012}. The advantage of such direct methods is that, unlike continuum-based models, they do not rely on phenomenological assumptions. Typically, the fluid-grain interactions are governed solely by a no-slip condition at the fluid-solid interfaces.
The limitations of these models come from the high computational cost, especially when applied to the solution of a nonlinear system of equations in three dimensions. The high costs are due to the need for a very small mesh size compared to the size of particles. Even if the LBM is known to minimize this computational cost, the minimal number of nodes necessary for accurate results in three dimensions still restricts the problem size to small numbers of particles \cite{Han2011}. In addition, it should be noted that incompressible flow represents a difficulty for the explicit methods such as conventional LBM-BGK, which are inherently based on density fluctuations. This feature makes the use of the explicit methods difficult in the situations of strong poromechanical coupling associated to incompressible or very weakly compressible phases.

To overcome the high computational cost of the microscale models, without introducing all the phenomenological assumptions of continuum-based methods, \textit{pore-network} (PN) models are a very attractive compromise \cite{Joekar2012}. PN models are based on a representation of the void space as a network of connected pores and throats, where the properties of the throats are supposed to reflect the effect of local void geometry on the flow. This modeling technique greatly reduces the number of unknowns as compared to micro-scale models. PN models have been extensively used for studying single-phase and multiphase flow in rigid porous materials, but they have been rarely coupled to models of deformable solid skeleton, although some examples exist in two dimensions for granular materials \cite{HakunoBook,Bonilla2004} and in three dimensions for fractured rocks \cite{Jing2001}.
A numerical model of the PN type has been developed recently for incompressible flow in sphere packings \cite{Char2011}. In this model, the spatial discretization leads to fluid elements whose sizes are of the same order as the size of the solid particles, thus introducing an intermediate scale between the microscale and the continuum scales mentioned above. At the same time it can reflect the effects of the local geometry of the pore space on the flow and gives predictive estimates of the permeability \cite{Tong2012}, unlike continuum based methods. Hereafter, we refer to this method as the Pore-scale Finite Volumes method (PFV). In the present paper, we present a coupling between the PFV method and a DEM model in three dimensions.

In the first part, the governing equations of the DEM and PFV models are recalled, and the coupling equations are established and discussed. The analogy of the system of equations with Biot's equations is highlighted in the particular case of incompressible phases. By such analogy, the DEM-PFV coupled model should be able to recover results of classical poroelasticity in boundary values problems, provided they share similar assumptions. This is the case for the simulation of an oedometer test, which is presented in the third part. In the fourth and last part, the coupling is applied to the simulation of a granular deposition problem. We will introduce microscale definitions of the stress and strain tensors for detailed analysis of the results. The evolution of stress, strain, and fluid pressure is presented and discussed. The onset of liquefaction events and fluid-solid transitions which are observed in the material behaviour, show the interest of following a discrete approach in simulating fluid-particle systems.

\section{Numerical model}
\subsection{Solid phase}
\subsubsection{Explicit DEM}
\label{DEM}
The discrete element method (DEM) has been extensively used to study soil and rock mechanics, providing, for instance, some insights into shear strength and deformation properties of geomaterials. 
The approach is fully micromechanical, the solid phase being modelled by defining the mechanical properties of the interaction between the grains that compose it \cite{Cundall1979}.
In what follows, we recall the generic aspects of the DEM method for the simulation of systems of spheres.
The algorithm presented uses the explicit finite difference scheme for time integration and assumes smooth contact behaviour. Those two aspects are sufficient requirements of our fluid-coupling algorithm. A large majority of publicly available softwares (including commercial and open-source) are using similar schemes according to our survey. It offers a good flexibility for incorporating additional physical effects and couplings (see e.g. \cite{Scholtes2009}). Hence, the coupling scheme we are developing can be potentially combined with a large number of existing DEM codes. Implementation details of the open-source code \textit{Yade} \cite{YadeDoc}, used for the present study, can be found in \v{S}milauer and Chareyre \cite{YADE-IMPL}.

\begin{figure}[t]
\centering
\includegraphics[width=6cm]{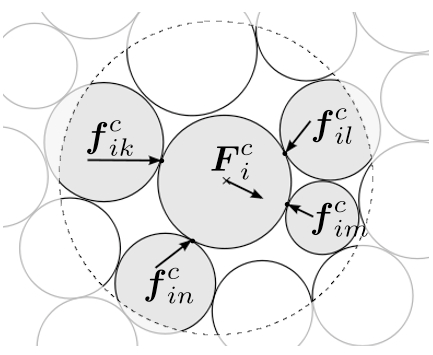}
\caption{Spherical particles interacting at contact points.}
\label{DEMgeom}
\end{figure}

We consider a system of $N$ spheres interacting with each other at contacts (fig. \ref{DEMgeom}). The kinematics of each sphere is described by six degrees of freedom (DOFs). We note $\mathbf{X}_i=\{\mathbf{x}_i,\mathbf{\theta}_i\}$ the \textit{generalized position} of sphere $i$, with $\mathbf{x}_i$ the position of its center of mass and $\mathbf{\theta}_i$ the rotation represented as a $\mathbb{R}^3$ vector (although in \textit{Yade} rotations are represented by quaternions, a rotation vector is mathematically equivalent and more convenient in the present derivation). The symbol $\mathbf{X}$ used without indice will refer to a vector containing all the DOFs of the system ($6\times N$) and will be called the global position vector. Similarly, $\mathbf{x}$ will contain all the translational DOFs.

The translational motion of each sphere in the system is governed by Newton's second law, which relates the forces exerted on a particle to its acceleration:
\begin{equation}
\label{single_newton}
m_i \mathbf{\ddot{x}_i}= \int_{\delta \Gamma_i}\sigma^s \mathbf{n} ds + \int_{\Gamma_i}\rho^s \mathbf{g} dv,
\end{equation}
where $m_i$ is the mass of a particle occupying the volume $\Gamma_i$, $\sigma^s\,\mathbf{n}$ is the stress applied at the particle surface in the direction of the unit normal $\mathbf{n}$, $\rho^s$ is the mass density and $\mathbf{g}$ is the gravitational acceleration. The first integral is the total force exerted on particle $i$ by the other particles, we note this force $\mathbf{F^c_i}$.

Assuming that the contact areas are negligible, the contact interactions can be represented by \textit{contact forces} $\mathbf{f_{ik}^c}$ acting at \textit{contact points}:  
\begin{equation}
\label{discrete_newton}
m_i \mathbf{\ddot{x}_i}= \sum_{k=0}^{n_c}{\mathbf{f_{ik}^c}} + m_i\,\mathbf{g}.
\end{equation}
where $n_c$ is the total number of contact points.

By introducing the global vectors $\mathbf{x}=\{\mathbf{x_i}\}$ and $\mathbf{F}^c=\{\mathbf{F^c_i}\}$ containing respectively the positions and forces for the $3\times N$ translational DOFs in the system, the evolution can be defined by the matrix relation: 
\begin{equation}
\label{newton}
\ddot{\mathbf{x}}=\mathbf{M}^{-1}(\mathbf{F}^c+\mathbf{W}),
\end{equation}
where $\mathbf{M}$ is the global mass matrix and the components of $\mathbf{W}$ contains the gravitational terms.  

A relation similar to eq.\ref{discrete_newton} holds for the rotational DOFs by replacing force and mass by the torque and inertia tensor, respectively. Hence, the system of eq.\ref{newton} must be supplemented with $3\times N$ equations for the rotational DOFs for completeness, in order to define completely the dynamics of the system: 
\begin{equation}
\label{newton_global}
\ddot{\mathbf{X}}=\mathbf{J}^{-1}(\mathbf{T}^c+\mathbf{W}'),
\end{equation}
with $\mathbf{J}$ the generalized inertia matrix, $\mathbf{T}^c$ the generalized force vector (including torques) ($\mathbf{W}'$ is simply $\mathbf{W}$ with zeros appended for the rotational DOFs).

The contact forces $\mathbf{f_{ik}^c}$ appearing in eq.\ref{discrete_newton} are computed according to contact laws, which often describe irreversible behaviour. For this reason, it is not possible in general to define a unique relation between positions and forces. Instead, it is usually only possible to define the rate of change of $\mathbf{f_{ik}^c}$ as a function of positions and their time derivatives:
\begin{equation}
\label{force_disp}
\mathbf{\dot{f}_{ik}^c}= b_{ik}(\mathbf{x_i},\mathbf{x_k},\mathbf{\dot{x}_i},\mathbf{\dot{x}_k})
\end{equation}
where the function $b_{ik}$ defines the constitutive behaviour of the contact between particles $i$ and $k$. Then, in terms of the global components, we can introduce an operator $B$ (nonlinear in general) corresponding to the summation of all forces and torques on the particles, so that
\begin{equation}
\label{claw}
\dot{\mathbf{F^c}}= B(\mathbf{X},\dot{\mathbf{X}})
\end{equation}

The explicit DEM method consists in integrating equations \ref{newton_global} and \ref{claw} with a time-stepping algorithm, updating positions and forces at each step.

The most common algorithm is based on a discretization of acceleration (left hand side of eq.\ref{newton}) with a centered second order finite difference scheme, which reads
\begin{equation}
\label{leapfrog}
\frac{\mathbf{X}_{t+\Delta t}-2 \mathbf{X}_{t}+\mathbf{X}_{t-\Delta t}}{\Delta t^2}=\mathbf{J}^{-1}\mathbf{T}^{c}_{t}.
\end{equation}
It results in an explicit equation for computing $\mathbf{X}_{t+\Delta t}$.

The contact forces are updated in the new configuration according to eq.\ref{force_disp}, where $\dot{\mathbf{F}}^c_{t+\Delta t/2}$ and $\dot{\mathbf{X}}_{t+\Delta t/2}$ are replaced by second order approximations at mid-step:
\begin{equation}
\label{force_rate}
\left\{ \begin{array}{ll}
\mathbf{F}^{c}_{t+\Delta t} = \mathbf{F}^{c}_{t} + B(\mathbf{X}_{t+\Delta t},\dot{\mathbf{X}}_{t+\Delta t/2}) \Delta t \\
\dot{\mathbf{X}}_{t+\Delta t/2} = \displaystyle  \frac{\mathbf{X}_{t+\Delta t}-\mathbf{X}_{t}}{\Delta t}
\end{array} \right.
\end{equation}

We may remark that this matrix representation of the problem is not very common in the literature on DEM. It is introduced here for an easier presentation and discussion of the coupled problem in the next sections. 

\subsubsection{Elastic-frictional contacts}
In the simulations that follow, the contacts will be modelized using an elastic-plastic contact law \cite{Cundall1979}. It does not imply any loss of genericity of the DEM-PFV coupling algorithm, which only depends on the explicit integration method formerly described.

The contact force is decomposed into its normal and tangential parts.
The normal force (compressive only) is proportional to the normal displacement $d_n$ and to the normal stiffness $k_n$,
\begin{equation}
f_n = \left\{ \begin{array}{ll} -k_n d_n \quad \textrm{if} \quad d_n\leq0\\
				0 \quad \textrm{if} \quad d_n > 0
	      \end{array} \right.
\end{equation}
$d_n$ is defined as $d_n=||\mathbf{x}_i-\mathbf{x}_j||-R_i-R_j$, where $R_i$ and $R_j$ are the radii of the particles in contact, so that $d_n=0$ when the spheres are exactly tangent.
The shear force depends on the shear stiffness $k_t$, and is integrated using eq.\ref{force_rate} where the relative velocity $\mathbf{\dot{d_t}}$ at contact point $\mathbf{x_c}$ depends on the spin $\mathbf{\omega}_{i,j}$ of each particle.
\begin{equation}
\left\{ \begin{array}{ll} \mathbf{\dot{f_t}} = -k_t \mathbf{\dot{d_t}}\\
			  \mathbf{\dot{d_t}} = \mathbf{\dot{x}}_j + \mathbf{\omega}_j \times (\mathbf{x_c}-\mathbf{x}_j) - \mathbf{\dot{x}}_i - \mathbf{\omega}_i \times (\mathbf{x_c}-\mathbf{x}_i)
	\end{array} \right.
\end{equation}

Coulomb's friction is introduced by imposing a maximum magnitude for $\mathbf{f_t}$:
\begin{equation}
  ||\mathbf{f_t}|| \leq tan(\phi) f_n
\end{equation}
where $\phi$ is the angle of Coulomb's friction at contacts.

$k_n$ and $k_t$ are defined, for each couple of particles $i$-$j$ in contact, as functions of an elastic modulus $E$ and a non-dimensional constant $a = k_t/k_n$:
\begin{equation}
\left\{ \begin{array}{ll} k_n = 2\,\frac{E\cdot R_i\cdot R_j}{(R_i+R_j)}\\ 
			  k_t = a \, k_n	\end{array} \right.
\end{equation}
This definition of stiffness is convenient for defining problems independently of the mean particles size, since it results in a constant ratio between $E$ and the effective bulk modulus for a given type of packing.  

\subsection{Fluid phase}
\label{sec:fluid}
\subsubsection{Viscous flow}
We assume that the porous medium is saturated by an incompressible fluid whose flow is governed at the micro-scale by Stokes equations, which express fluid mass and moment conservation at small Reynolds and large Stokes numbers.
Situations in which the Reynolds number is small are called \textit{viscous flows}, because viscous forces arising from shearing motions of the fluid predominate over inertial forces associated with acceleration or deceleration of fluid particles. The Reynolds number is defined as:
\begin{equation}
 \label{Renumb_def}
 Re = \frac{\rho^f \mathbf{u} \cdot \nabla \mathbf{u}}{\mu \nabla^2 \mathbf{u}}
\end{equation}
where $\rho^f$ is the fluid density, $\mathbf{u}$ its velocity, $\mu$ its viscosity. By noting as $U$ a characteristic fluid velocity for the problem and $d$ a characteristic length describing the problem (e.g. the average throat dimension), the Reynolds number can be computed as,
\begin{equation}
 \label{Renumb}
 Re = \simeq \frac{\rho^f\,U\,d}{\mu}
\end{equation}
\\ In fluid-particle systems, the Stokes number is a dimensionless parameter that is defined as the ratio between a viscous diffusion term and a term related to the fluid acceleration:
\begin{equation}
\label{stknumb1}
 S_t = \frac{\mu \frac{\partial^2 u}{\partial x^2}}{\rho^f \frac{\partial u}{\partial t}}
\end{equation}
Using the same notation of eq.\ref{Renumb}, and noting as $\tau_c$ a characteristic time of rearrangement of the particles, Stokes number can be written as:
\begin{equation}
\label{stknumb2}
 S_t = \frac{\tau_c}{\tau_{\nu}} = \frac{\mu \tau_c}{d^2 \rho^f}
\end{equation}
The motion of the particles produces in fact a poral flow whose relaxation time is of the order of magnitude of $\tau_{\nu} = d^2/\nu$ (remember that $\nu=\mu/\rho^f$). A situation in which $\tau_{\nu}<<\tau_c$ means that the flow induced by the particles displacement attains rapidly a new equilibrium. In this case ($S_t>>1$), the acceleration of the fluid can be neglected and the hypothesis of steady laminar flow holds.

For small Reynolds and large Stokes numbers, Stokes equations, for the description of the flow, read:
\vskip-.6cm
\begin{eqnarray}
\label{stokes1}
\triangledown p = \mu \triangledown^2 \mathbf{u} \\
\triangledown \cdot \mathbf{u} = 0
\end{eqnarray}
where $\mathbf{u}$, and $p$ are the microscopic fluid velocity and piezometric pressure, respectively. The piezometric pressure $p$ is related to the absolute pressure $p^a$ via $p = p^a - \rho^f gz$, with $g$ the gravitational acceleration and $z$ the depth coordinate. A no-slip condition is assumed for the fluid velocity at the grain boundaries.

Doing the hypothesis of slow viscous flow, the flow and the forces exerted on the particles can be computed using the PFV model recently developed \cite{Char2011}. The key aspects of this method are summarized below.

\subsubsection{The PFV model}
\begin{figure}[t]
\centering
\includegraphics[width=12cm]{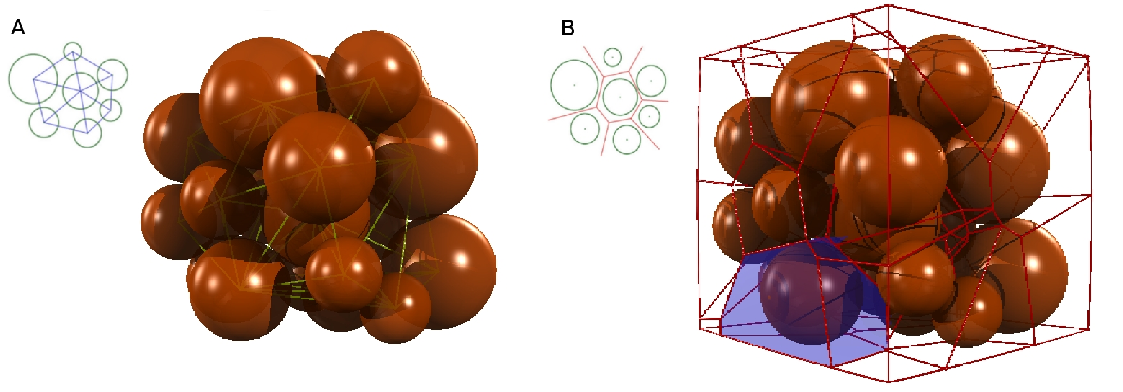}
\caption{Regular triangulation (A) and its dual Voronoi diagram (B).}
\label{Trg_Tess}
\end{figure}

\begin{figure}[t]
\centering
\includegraphics[width=12cm]{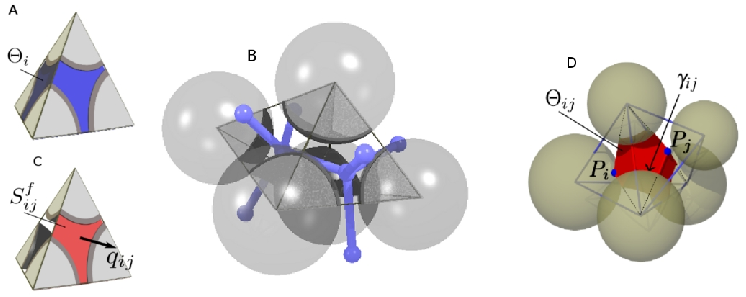}
\caption{Volume of fluid in a pore (A), adjacent pores and local connections (B), fluid domain of pore contour (C), pore partition for hydraulic radius definition (D) \cite{Char2011}}
\label{PoreNetwork}
\end{figure}

Regular triangulation and its dual Voronoi graph are used to discretize the void space. A system of tetrahedra arises from the triangulation in a 3D framework, each tethraedron representing a pore (see fig.\ref{Trg_Tess}(A) and fig.\ref{PoreNetwork}). The vertices of the triangulations are the spheres of the packing, so that the displacement of the particles will be reflected by the deformations of the mesh elements. The dual Voronoi diagram constitutes a network whose edges do not cross the solid phase (see fig.\ref{Trg_Tess}(B)). Such network represents the flow path within the porous sample and allows the formulation and resolution of the flow problem, practically upscaling Stokes equations at the pores scale. The pressure field is defined by the values of pressure in each tetrahedral element, located at the vertices of the Voronoi's graph. 
The mass balance equation integrated on the pore $i$, and recast into a surface integral by using the divergence theorem, gives:
\begin{eqnarray}
\label{continuity}
\dot{V_i^f} = \int_{S_{ij}} {(\mathbf{u} - \mathbf{v}) \cdot \mathbf{n} ds} = \sum_{j=j_{1}}^{j_{4}} q_{ij}
\end{eqnarray}
where $V_i^f$ is the total pore volume, $(\mathbf{u}-\mathbf{v})$ is the velocity of the fluid relative to that of the solid phase. The integral on facet $S_{ij}$ gives a flux exchanged between adjacent tetrahedra, noted $q_{ij}$ (see fig.\ref{PoreNetwork}C). The sum of fluxes equals the rate of volume change of the pore $\dot{V_i^f}$, which is related to the velocity of the particles. This sum can be seen as a discrete divergence operator applied on fluid velocity.
\\ \\
A key aspect of the model is the expression of the flux $q_{ij}$ through a facet as a function of the local geometry and pressures in the pores. Stokes equations imply a linear relation between pressure gradients and fluxes. The expression can therefore take the form
\vskip-.6cm
\begin{eqnarray}
\label{fluxes}
q_{ij} = g_{ij} \frac{p_i - p_j}{l_{ij}}
\end{eqnarray}
where a local pressure gradient is defined as the ratio between the pressure drop $p_i - p_j$ and the distance $l_{ij}$ between pores $i$ and $j$ (euclidean distance between the Voronoi vertices associated to each pore). $g_{ij}$ is a term expressing the hydraulic conductance of the domain $\Theta_{ij}$ represented on fig.\ref{PoreNetwork}D.
Combining equations \ref{continuity} and \ref{fluxes} gives a relation linking the discrete pressure field to the velocities of the particles:
\begin{eqnarray}
\label{poisson}
\dot{V_i^f} =  \sum_{j=j_{1}}^{j_{4}} \frac{g_{ij}}{l_{ij}}(p_i-p_j)
\end{eqnarray}
The following definition has been proposed and validated for $g_{ij}$ \cite{Char2011,Tong2012}:
\vskip-.6cm
\begin{eqnarray}
\label{loconduct}
g_{ij} = \frac{S_{ij}^f {R_{ij}^h}^2}{2\mu}
\end{eqnarray}
where $R_{ij}^h$ is the hydraulic radius (defined hereafter), $S_{ij}^f$ is the area occupied by the fluid in facet $S_{ij}$, and $\mu$ is the fluid viscosity. 
The hydraulic radius was defined as the ratio between the volume $\Theta_{ij}$ occupied by the fluid and the area $\gamma_{ij}$ of solid-fluid interface (see fig.\ref{PoreNetwork}D).

\subsubsection{Forces}
The total force $\mathbf{F^{f}_{i}}$ exerted by the fluid on a particle $i$ results from the pressure and viscous stress acting at the surface:
\vskip-.6cm
\begin{eqnarray}
\mathbf{F^{f}_{i}} = \int_{\partial \varGamma_i} (-p^{a} \mathbf{n} + \boldsymbol\tau \mathbf{n}) ds
\end{eqnarray}
where $\partial \varGamma_i$ denotes the solid surface of the particle $i$, $p^{a}$ the absolute pressure and $\boldsymbol\tau$ the viscous shear stress tensor. Remembering that $p = p^a - \rho^f gz$, $\mathbf{F^{f}_{i}}$ can be evaluated by summing three terms (\cite{Char2011}):
\vskip-.6cm
\begin{eqnarray}
\label{fluid_forces}
\mathbf{F^{f}_{i}} = \int_{\partial \varGamma_i} -\rho^f g z \, \mathbf{n} ds + \int_{\partial \varGamma_i} p\, \mathbf{n} ds + \int_{\partial \varGamma_i} \boldsymbol \tau \, \mathbf{n} ds
\end{eqnarray}
where the first term denotes the buoyancy force, the second term the integral of piezometric pressure, and the third term the integral of the viscous stress. The second and third term vanish if the fluid is at hydrostatic equilibrium (constant $p$).

\subsection{Hydromechanical coupling}
The previous studies on the PFV method were limited to flow in rigid assemblies of spheres. Therefore, the hydromechanical coupling was not considered. In this section, we detail the micro-scale coupling equations that appear when the DEM and PFV models are combined in a unified framework and we discuss the relation with the field equations of classical poroelasticity.

\label{sec:coupling}
\subsubsection{Coupling equations}
The coupling is defined by two matricial relations. The first one corresponds to the mass conservation equation \ref{poisson}, written for all pores, complemented with the boundary conditions:
\begin{equation}
\label{flow_solution2}
\mathbf{G} \mathbf{P} = \mathbf{E}\,\dot{\mathbf{x}}+ \mathbf{Q}_q+\mathbf{Q}_p,
\end{equation}
where $\mathbf{G}$ is the conductivity matrix containing terms $g_{ij}/l_{ij}$ of equation \ref{poisson}, $\mathbf{P}$ the column vector containing all values of pressure, and $\mathbf{E}$ is the matrix defining the rates of volume change such that $\dot{V_i^f}= (\mathbf{E}\,\dot{\mathbf{x}})_i$. $\mathbf{Q}_q$ and $\mathbf{Q}_p$ are flux vectors reflecting boundary conditions, respectively source terms (imposed fluxes) in $\mathbf{Q}_q$ and imposed pressures in $\mathbf{Q}_p$.
Solving this equation gives the field of fluid pressure $\mathbf{P}$ as function of particles velocity $\dot{\mathbf{x}}$.

The second relation results from the addition of the fluid forces to the contact forces in Newton's equation \ref{newton}, which becomes:
\begin{equation}
\label{brute_newton_coupled}
\mathbf{M}\ddot{\mathbf{x}}=\mathbf{F}^c+\mathbf{W}+\mathbf{F}^f,
\end{equation}
where $\mathbf{F}^c$ and $\mathbf{F}^f$ are the contributions of contact forces (as they appear in eq.\ref{newton}) and fluid forces (eq.\ref{fluid_forces}), respectively.

The expression of $\mathbf{F}^f$ of eq.\ref{fluid_forces}, can be expressed in matricial form as a function of the pressure field $\mathbf{P}$ and a matrix $\mathbf{S}$ reflecting the local geometry of the sphere packing (please refer to \cite{Char2011} for details):
\begin{equation}
\label{fluidforces}
\mathbf{F}^f=\mathbf{S}\mathbf{P}
\end{equation}

\subsubsection{Coupled problem}
Combining equations \ref{flow_solution2} and \ref{brute_newton_coupled}, we end up with an explicit ordinary differential equation of order 2, where $\mathbf{X}$ is the only remaining unknown:
\begin{equation}
\label{coupled}
\ddot{\mathbf{X}}=\mathbf{M}^{-1}(\mathbf{F}^c+\mathbf{W}+\mathbf{S} \mathbf{G}^{-1} (\mathbf{E} \cdot \dot{\mathbf{X}} + \mathbf{Q}_q + \mathbf{Q}_p)).
\end{equation}

Integrating this equation numerically poses no major theoretical difficulty. The methods used in finite element solvers may also be used here (see e.g. \cite{Detournay1993}). However, attention must be paid to the method choosen in order to minimize the computational cost. For this purpose, we developed a semi-implicit scheme \cite{CatalanoPhD,Catalano2011a} which makes the resolution of the coupled problem possible with quite acceptable computation times. The overhead of running a coupled simulation with this scheme is of the order of 100\% of the computation time of the same simulation without fluid.

We note that the constitutive relations will to lead to energy dissipation in the simulated systems by either frictional effects at contacts (eq. 11) or viscous effects due to the fluid (eq. 22). It is relatively common in DEM to introduce additional sources of dissipation, damping the equation of motion (eq. \ref{discrete_newton}) artifially \cite{Chareyre2005} for a faster convergence to static equilibrium. Such numerical damping is not required in the DEM-PFV since it is naturally damped, and it would in fact lead to erroneous results regarding the time evolution of the systems. In what follows, the results are obtained without introducing any form of numerical damping.

\subsection{Relation with classical poromechanics}
Although the coupling equations were obtained only from micro-scale considerations, we will show that they may be seen as a discrete form of the field equations of conventional Biot's theory of poroelasticity for quasi-static deformations.
It is worth noting this feature since the next section presents the comparisons with Terzaghi's analytical solution of the monodimensional consolidation problem. \\
In the case of incompressible phases (Biot’s coefficient = 1), the coupling equations of Biot's theory are the Poisson's equation that result from the continuity of Darcy's velocity \cite{Detournay1993} :
\begin{gather}
\label{biot2}
\frac{\partial}{\partial t}(\nabla\cdot \mathbf{dx}) + \nabla\cdot(-K_d\,\nabla p) = 0,
\end{gather}
and the equation of local equilibrium:
\begin{gather}
\label{biot1}
-\nabla\cdot\boldsymbol{\sigma'} + \nabla p  = (1-n)(\rho^s-\rho^f)\mathbf{g},
\end{gather}
where $p=p^a-\rho^f \,g\, z$ is the excess (''piezometric'') pore pressure and $K_d$ is the hydraulic conductivity of the medium. $\sigma'$ represents Terzaghi's effective stress and $\mathbf{dx}$ is the displacement of the solid phase.

In small strain linear elasticity, $\boldsymbol{\sigma'}=\mathbf{C}\frac{1}{2}(\nabla \mathbf{dx} + \nabla \mathbf{dx}^\mathrm{T})$ with $\mathbf{C}$ the stiffness tensor. Substituting $\boldsymbol{\sigma'}$ by this expression in eq.\ref{biot1} gives a Navier-type equation where $\nabla p$ can be seen as a body force:
\begin{gather}
\label{biot3}
-\nabla\cdot\mathbf{C}\frac{1}{2}(\nabla \mathbf{dx} + \nabla \mathbf{dx}^\mathrm{T}) + \nabla p  = (1-n)(\rho^s-\rho^f)\mathbf{g}.
\end{gather}

In the light of this system of partial differential equations, we can reconsider the equations of the pore-scale DEM-PFV formulation. Firstly, we observe that in the case of steady rates of deformation, $\ddot{\mathbf{X}}$ vanishes and eq.\ref{coupled} becomes an equilibrium equation. Secondly, we note that in the special case when all contacts behave purely elastically, a linear relation exists between the contact force vector $\mathbf{F}^c$ and the generalized displacement $\mathbf{dX}$ via a global stiffness matrix $\mathbf{C}^s$ \cite{Agnolin2007}. Hence, the system satisfies a relation of the form:
\begin{equation}
\label{equilibrium}
\mathbf{C}^s\,\mathbf{dX} +\mathbf{W}+\mathbf{S} \mathbf{P}=0
\end{equation}
Considering only one particle $i$ of the system, interacting with $n_c$ particles in contact, and with $n_f$ incident fluid cells, eq.\ref{equilibrium} implies
\begin{equation}
\label{equilibrium_particle}
\sum_{k=0}^{k=n_c}{\mathbf{C^s_{ik}}(\mathbf{dX}_i-\mathbf{dX}_k)} + V_i\,\rho^s\,\mathbf{g} + \sum_{k=0}^{k=n_f}{{S}_{ik} p_k} =0 .
\end{equation}
In this equation $\mathbf{C}^s_{ik}$ is the so called rigidity matrix \cite{Agnolin2007}, that is multiplied by the relative displacements $\mathbf{dX}_i-\mathbf{dX}_k$ to give the contact forces; the second term expresses the weigth of particle $i$ ($V_i$ = particle volume); the last term sums the contributions of fluid pressure in incident pores. Equation \ref{equilibrium_particle} can be seen as the discrete form of a Navier equation, where $\sum_{k=0}^{k=n_c}{\mathbf{C}^s_{\cdot k}(\cdot - \mathbf{dX}_k)}$ is an operator defined for the discrete displacement field, and equivalent to the operator $\int_V \nabla \cdot {\mathbf{C}}(\nabla \cdot + \nabla^T \cdot)$ that would arise in the conventional finite volume formulation for a continuum. Similarly, the third term is a discrete operator equivalent to the integral of the pressure gradient $\int_V\nabla\cdot$. We note that the microscale counterpart of the divergence of the effective stress is a sum of contact forces.

The analogy between eq.\ref{poisson} and eq.\ref{biot2} is direct since in the former, by definition, the two terms represent the local rate of volume change of the pore space and the divergence of the fluid velocity averaged in the pore.

Consistently, we also remark that the assembled matrices for the coupled boundary value problem, as given by equations \ref{flow_solution2} and \ref{equilibrium}, do not differ from the ones obtained through the discretization of Biot's equations using the FVM \cite{Naumovich2006} or FEM \cite{Detournay1993} methods. As a last note, we can remark that the sets of elements that we obtain from the triangulation and the tessellation systems are also quite similar to the ones found in unstructured FV mesh \cite{Mahesh2004,Norris2011}.

From this comparison, we can conclude that the DEM-PFV coupled model may recover results of classical poroelasticity in boundary values problems, provided they share similar assumptions (this is further discussed in the next section). The 1D diffusion problem known as Terzaghi's consolidation is well suited for such comparison. It is used as a benchmark test in the next section for the validation of the model.

\section{Oedometer test}
\label{consolid} 
The consolidation process is a classical hydro-mechanical problem, of primary importance in geomechanics. In the case of incompressible phases and when the deformation occurs in only one direction, as in oedometer tests (fig. \ref{BC_Oedometer}), Biot's theory of poroelasticity coincides with the classical solution of Terzaghi.

The equation of monodimensional consolidation is a diffusion equation on $p$, that reads:
\vskip-.6cm
\begin{eqnarray}
\label{anal_consolidation}
\frac{\partial p}{\partial t} = C_v \frac{\partial^2 p}{\partial z^2}
\end{eqnarray}
where $z$ is the coordinate along the axis of application of the load (vertical direction), $C_v$ the consolidation coefficient, defined as follows:
\vskip-.6cm
\begin{eqnarray}
\label{consol_coeff}
C_v = \frac{K_d\, E_{oed}}{\rho^f \, g} \quad \left[\frac{m^2}{s}\right]
\end{eqnarray}
where $K_d$ is the coefficient of hydraulic conductivity of the soil, $E_{oed} = \Delta \sigma^{'}_v/\Delta \varepsilon_v$ the oedometric modulus.
A non-dimensional time parameter is introduced, $T_v$, defined as:
\vskip-.6cm
\begin{eqnarray}
\label{Terzaghi_solution_2}
T_v = \frac{C_v t}{L_d^2}
\end{eqnarray}
where $t$ is the time, and $L_d$ the longest drainage path for the generic fluid particle.

By noting $Z=z/H$, with $H$ the height of the sample, the analytical solution of the problem of eq.\ref{anal_consolidation} (Taylor, 1948), expressing the evolution of the interstitial pressure $p$, reads:
\begin{equation}
 \label{Taylor}
 p(Z,T_v) = \sum^{\infty}_{m=0} \frac{4 p_0}{\pi \cdot (2m+1)}\cdot sin(\frac{\pi}{2} \cdot (2m+1) \cdot Z)\cdot e^{-(\frac{\pi}{2} \cdot (2m+1))^2 \cdot T_v}
\end{equation}

As the interstitial pressure dissipates, the deformation of the sample takes place. The \textit{average degree of consolidation} expresses the evolution of the total settlement over the final settlement that would result at the end of the consolidation process, and is defined as:
\begin{equation}
 \label{anal_settlement}
 \frac{s(t)}{s_{final}} = 1-\sum_{m=0}^{\infty} \frac{8}{(\pi \cdot (2m+1))^2} \cdot e^{-(\frac{\pi}{2} \cdot (2m+1))^2 \cdot T_v}
\end{equation}

\subsection{Numerical simulations}
\label{num_results}
The 1D consolidation problem will be used as a benchmark test for the DEM-PFV coupling. The numerical solution will be compared directly to the analytical solution of eq.\ref{Taylor} and eq.\ref{anal_settlement}. The DEM-PFV model is consistent with Terzaghi's assumptions under the following conditions:
\begin{itemize}
 \item c1) The deformation is small and the stress-strain behaviour is linear;
 \item c2) Biot's coefficient = 1 (incompressible phases);
 \item c3) The displacement is unidirectional;  
 \item c4) The permeability and the elastic properties are constant in space and time.
\end{itemize}
 
The c1) condition can be obtained with the DEM for small deformations, as shown below. The c2) hypothesis holds in the DEM-PFV model since both phases are incompressible. Boundary conditions can be specified to be consistent with the c3) condition, but the displacement inside the material will not be purely unidirectional due to local fluctuations in the displacement field. By definition, a discrete model generally does not satisfy the c4) condition, since the heterogeneous arrangement of particles leads to fabric heterogeneities, then to a spatial variability of the elastic properties and the local permeabilities. In addition, the evolution of the microstructure during the deformation may induce an evolution of the mechanical and hydraulic properties of the medium. 
\\ In a first simulation, however, the permeability and the oedometric modulus will be nearly constant in time since the deformation will be very small. In a second simulation, conditions c1) and c4) conditions will be relaxed and the consequences will be commented.

\begin{figure}[t]
\centering
\includegraphics[width=12cm]{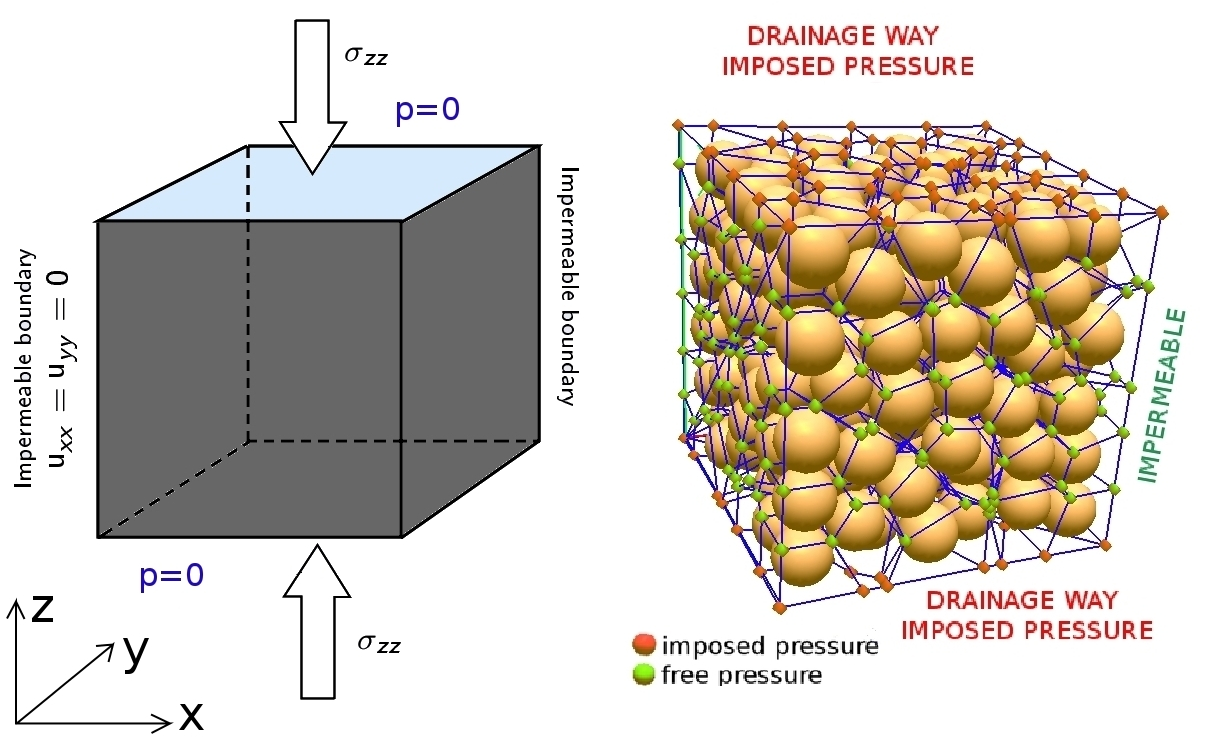}
\caption{Oedometric boundary conditions and DEM model of oedometer test (with less particles than in the actual simulation).}
\label{BC_Oedometer}
\end{figure}
\begin{figure}[t]
\centering
\includegraphics[width=8cm]{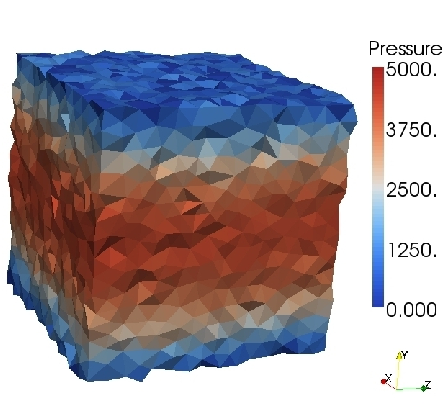}
\caption{Pressure field at $T_v=0.1$.}
\label{BC_Oedometer_2}
\end{figure}

The oedometer test was simulated on a cubic sample, sized $0.1m$, bounded by rigid plates. The boundary conditions are shown on figure \ref{BC_Oedometer}. Lateral displacements are prevented ($u_{xx} = u_{yy} = 0$). The fluid pressure $p=0\,kPa$ is imposed at $z=0$ and $z=H$ (corresponding to a two-way drainage with $L_d=H/2$ in eq.\ref{Terzaghi_solution_2}). $5000$ slightly polydispersed grains were employed to build the sample. They were first compacted isotropically using the REFD growth algorithm \cite{Chareyre2002}, by which a confinement stress of $5$ kPa is applied. A relatively dense sample was created ($n\simeq0.36$), to minimize the dispersion of pores' dimension and avoid strong spatial heterogeneities within the sample. Then, the oedometer test itself was simulated by applying an increment of stress $\Delta\sigma_{ext}=5$ kPa on the top plate. Table \ref{oedometer} gives the main parameters of the simulation. 

\begin{figure}
 \centering
\includegraphics[width=14cm]{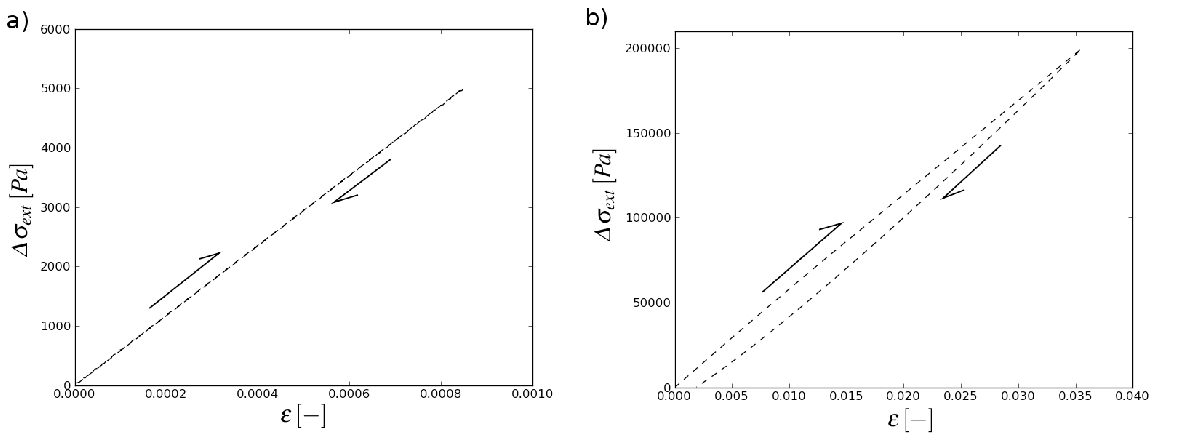}
\caption{Simulated load-unload cycle on the dry sample with $\Delta \sigma_{ext} = 5\,kPa$ (a) and $\Delta \sigma_{ext} = 200\,kPa$ (b).}
\label{LoadUnload}
\end{figure}

In addition to the oedometer test, two independent simulations were performed on the same sample. First, a compression test on the same sample but in dry conditions, with the same loading path as the oedometer test, was used to evaluate the oedometric modulus of the sample. In a loading-unloading cycle at $\Delta\sigma_{ext}=5\,kPa$, the response is completely reversible, showing that we are in the purely elastic regime at such stress-strain level (see fig.\ref{LoadUnload}(a)). The final strain obtained with the dry sample ($8.525 \cdot 10^{-4}$) defines an oedometric modulus $E_{oed} = 5895\,kPa$. Second, a permeability test was simulated on the initial geometry by imposing $p=0$ at $z=0$ and $p=1$ at $z=H$. The solution of this simulation gave fluxes that were used to determine the equivalent permeability of the sample. The permeability obtained from the permeameter simulation is $K_d=7.07623\cdot 10^{-5} \, m/s$.

\begin{table}
\caption{Oedometer test - Input data of the test of fig.s \ref{Comparison_1},\ref{Comparison_2}.}
\label{oedometer}
\centering
\begin{tabular}{*{3}{c}}
\toprule
& INPUT DATA & \\
\midrule
Number of grains & [-] & 5000 \\
$\rho^f$ & [kg/m$^3$] & 1000 \\
$\rho^s$ & [kg/m$^3$] & 2600 \\
Sample dimensions & [m] & 0.66 x 0.66 x 0.66 \\
$\mu$ & [kPa $\cdot$ s] & 0.25 \\
$d_{50}$ & [m] & 0.0395 \\
Confinement stress & [kPa] & 5 \\
$\Delta \sigma_{ext}$ & [kPa] & 5 \\
$p_0$ & [kPa] & 0 \\
$E$ & [kPa] & 15000 \\
$k_t/k_n$ & [-] & 0.5 \\
\\
\bottomrule
\end{tabular}
\end{table}

\begin{figure}
\centering
\includegraphics[width=10cm]{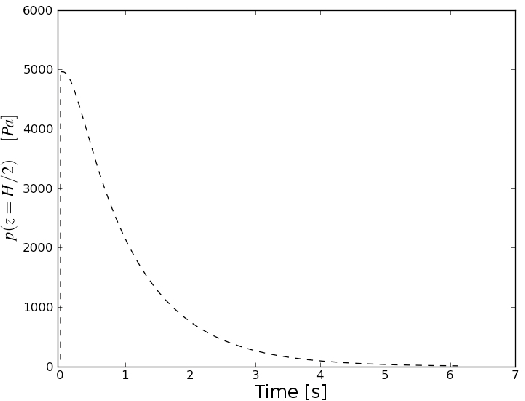}
\caption{Oedometer test - Fluid pressure measured at $z=H/2$ during the consolidation process.}
\label{Comparison_1}
\end{figure}

\begin{figure}
\centering
\includegraphics[width=14cm]{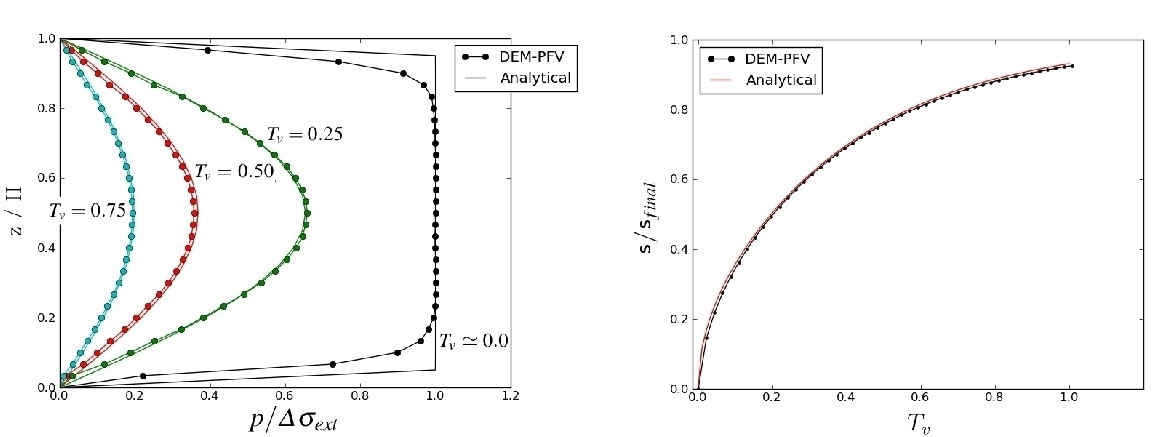}
\caption{Oedometer test - Evolution of pore pressure ($\Delta \sigma_{ext}=5\,kPa$) (left) and settlement (right)}
\label{Comparison_2}
\end{figure}

The numerical results that were obtained for the oedometer test are summarized in table \ref{resultsEDO1}. Fig.\ref{Comparison_1} shows the evolution of excess pore pressure at half the total height of the sample. The plotted value is an average computed on the plane $(x,y,z=H/2)$. Fig.\ref{Comparison_2} (left) shows the evolution of pore pressure in space ($z/H$) and time ($T_v$). The pore pressure is normalized by the increment of applied stress $\Delta \sigma_{ext}$. Fig.\ref{Comparison_2} (right) shows the evolution of the settlement $s$, normalized to the value $s_{final}$ obtained on the dry sample.

The excess pore pressure (plotted on fig.\ref{Comparison_1}) rose up almost instantaneously to $5\,kPa$ (= $\sigma_{ext}$ = $p_{max}$) and then gradually decreased. We remind that in Terzaghi's formulation, the initial condition is $p = \sigma_{ext}$, at $t=0$ $(T_v=0)$. In our simulation, $p = 0$ initially, and the maximum value was obtained in finite time: $p = \sigma_{ext}$ at $T_v=3.2\cdot10^{-5} \simeq 0$. This short time lag can be explained by the inertia of the system: pore pressure results from the velocity of the solid phase, which is not established instantaneously. This short delay is also found in experiments \cite{Gunaratne1996} and in other numerical simulations \cite{Boutt2007,Chen2007}. For the rest of the process, the evolution of pressure in space and time is found to be in good agreement with the analytical solution, as it can be seen on fig.\ref{Comparison_2} (left). The same conclusion holds for the evolution of the settlement, as shown on fig.\ref{Comparison_2} (right).

\begin{table}
\caption{Oedometer test - Numerical result (see fig.s \ref{Comparison_1},\ref{Comparison_2}).}
\label{resultsEDO1}
\centering
\begin{tabular}{*{3}{c}}
\toprule
& RESULTS & \\
\midrule
Final strain & [-] & $8.525 \cdot 10^{-4}$ \\
$E_{oed}$ & [kPa] & 5895 \\
$K_d$ & [m/s] & $7.07623 \cdot 10^{-5}$ \\
\\
\bottomrule
\end{tabular}
\end{table}


\subsection{Microscopic stress and strain}
\begin{figure}[t]
\centering
\includegraphics[scale=0.4]{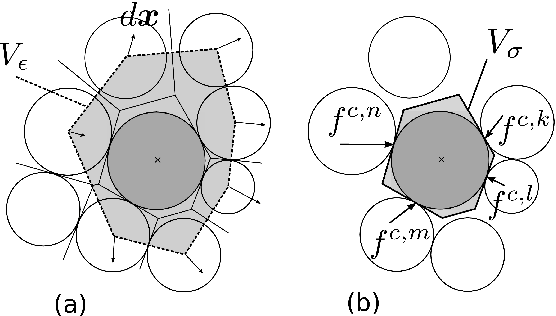}
\caption{Particle-centered domains for the definition of micro-strain (a) and micro-stress (b).}
\label{micro-domains}
\end{figure}

For a more detailed analysis of the results, the computation of microscale strain and stress tensors has been implemented in Yade-DEM code \cite{YadeDoc}. The tensors are defined in particle-centered volumes $V_\epsilon$ and $V_\sigma$ (fig. \ref{micro-domains}).
The microscopic stress associated to one particle is defined as a sum over the contacts \cite{drescher1972,Bagi1996}:
\begin{equation}
 \overline{ \boldsymbol \sigma} = \frac{1}{V_\sigma} \sum_k \mathbf{x}^{c,k} \otimes\mathbf{f}^{c,k}
\end{equation}
where $\mathbf{x}^{c,k}$ is a contact point and $\mathbf{f}^{c,k}$ the corresponding force. $V_\sigma$ is the reference volume associated to the particle in the Voronoi tesselation. Note that this tensor does not reflect the average stress in the solid material. For this purpose, we would have to divide by the volume of the particle, not by $V_\sigma$, and the stress applied by the fluid on the contour should be accounted for. Instead, this micro-stress only reflects that part of the external loading that acts through the contact network. For this reason, it can be seen as the microscale analogue of Terzaghi's effective stress, as will be confirmed by the results.

The microscale strain tensor for one particle is defined as a function of the displacements of the particles adjacent to that particle in the regular triangulation, which define the polyhedral domain $V_\epsilon$. The average displacement gradient in an equivalent continuum that would be contained in $V_\epsilon$ is
\begin{equation}
 <\nabla \mathbf{dx}> = \frac{1}{V_\epsilon} \int_{V_\epsilon} \nabla \mathbf{dx}  dv = \frac{1}{V_\epsilon} \int_{\partial V_\epsilon} \mathbf{dx}\otimes\mathbf{n} ds,
\end{equation}
where the displacement $\mathbf{dX}$ on the contour $\partial V_{\epsilon}$ is defined as a piecewise linear function, equal to the displacement of the particles at the vertices and linear on each facet. This expression generalizes in three dimension the expression proposed in \cite{Cundall1982}. The micro-strain is then obtained as the symmetric part of the gradient:
\begin{equation}
 \overline{\boldsymbol \varepsilon} = \frac{1}{2} (<\nabla \mathbf{dx}> + <\nabla\mathbf{dx}>^T)
\end{equation}

Both tensors may not be very meaningful at the scale of one particle alone, and there is no clear constitutive relation between $\overline{\boldsymbol \varepsilon}$ and $\overline{ \boldsymbol \sigma}$ at such a small scale \cite{Bagi1996}. They have however the interesting property to converge, respectively, to the average strain and stress in an equivalent continuum when they are averaged in larger domains containing many particles \cite{Calvetti1997}. Hence, we can define stress and strain at a certain height $z$ in the sample as weighted averages of the tensors associated to the particles present at (or near) this height. It let us plot profiles of stress and strain as functions of $z$.

\begin{figure}
\centering
\includegraphics[width=14cm]{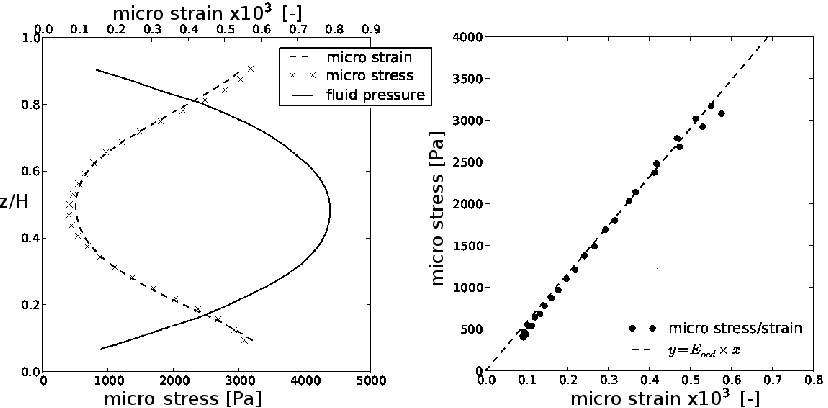}
\caption{Consolidation problem - $T_v=0.10$. On the left, profile of fluid pressure $p$, micro-stress $\overline{ \boldsymbol \sigma}_{zz}$ and micro strain $\overline{\boldsymbol \varepsilon}_{zz}$. On the right, the same data are plotted on a $\overline{ \boldsymbol \sigma}_{zz}$ vs. $\overline{\boldsymbol \varepsilon}_{zz}$ graph.}
\label{StressStrain}
\end{figure}

\begin{figure}
\centering
\includegraphics[width=9cm]{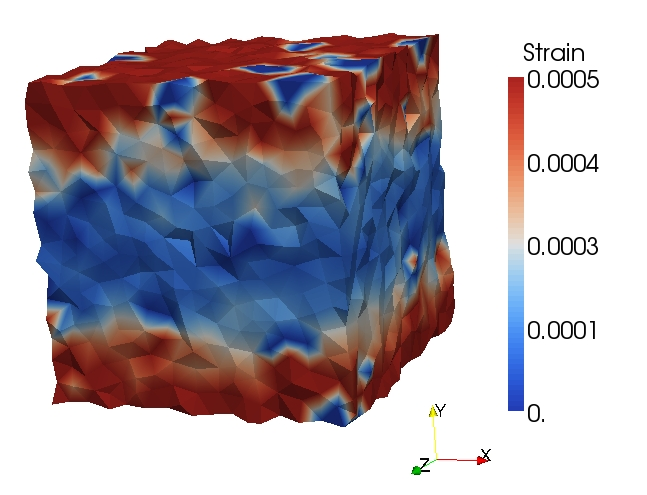}
\caption{Strain field at $T_v=0.10$ - 3D-visualization.}
\label{deformation}
\end{figure}

So, fig.\ref{StressStrain} (left) shows the profiles of fluid pressure $p$, micro-strain $\overline{\boldsymbol \varepsilon}_{zz}$ and micro-stress $\overline{ \boldsymbol \sigma}_{zz}$ at $T_v \simeq 0.10$. As it was expected, fluid pressure and effective stress profiles are complementary, as stated by the Terzaghi's effective stress principle $\sigma = \sigma'+p$. The linear relation between the micro-stress and the micro-strain can be observed in the right diagram (fig.\ref{StressStrain} (right)).
Fig.\ref{deformation} shows a 3D-visualization of the deformation field within the sample, at the same time $T_v \simeq 0.10$. A profile similar to the ones of fig.\ref{BC_Oedometer_2} can be observed, although the strain field reflects more local fluctuations than the pressure field.

\subsection{Nonlinear consolidation problem}
\label{nonlinearCons}
In this section, the influence of conditions c1) and c4) (see previous section) on the final solution is examined by simulating larger deformations. The input data are the same than the ones summarized in table \ref{oedometer}, except that the amplitude of the stress increment is larger: $\Delta\sigma_{ext}=200$kPa.
The final strain obtained through the dry compression test was $3.94 \cdot 10^{-2}$. On fig.\ref{LoadUnload}(right) it can be seen how the initial state was not recovered after unloading the sample. Table \ref{resultsEDO2} summarizes the results for this new simulation. The initial and secant values of the permeability, and the intitial and secant values of the oedometric modulus are reported, as well as the respective consolidation coefficients, computed using eq.\ref{consol_coeff}.

\begin{table}
\caption{Nonlinear oedometric consolidation - Numerical result.}
\label{resultsEDO2}
\centering
\begin{tabular}{*{3}{c}}
\toprule
& RESULTS & \\
\midrule
Final strain & [-] & $3.94 \cdot 10^{-2}$ \\
Initial Permeability & [m/s] & $7.07623e-05$ \\
Secant Permeability & [m/s] &  $6.41317e-05$ \\
Initial $E_{oed}$ & [kPa] & $5895$ \\
Secant $E_{oed}$ & [kPa] & $5620$ \\
Initial $Cv$ & [-] & $0.0417$ \\
Secant $Cv$ & [-] & $0.0360$ \\
\\
\bottomrule
\end{tabular}
\end{table}

\begin{figure}
 \centering
\includegraphics[width=140mm]{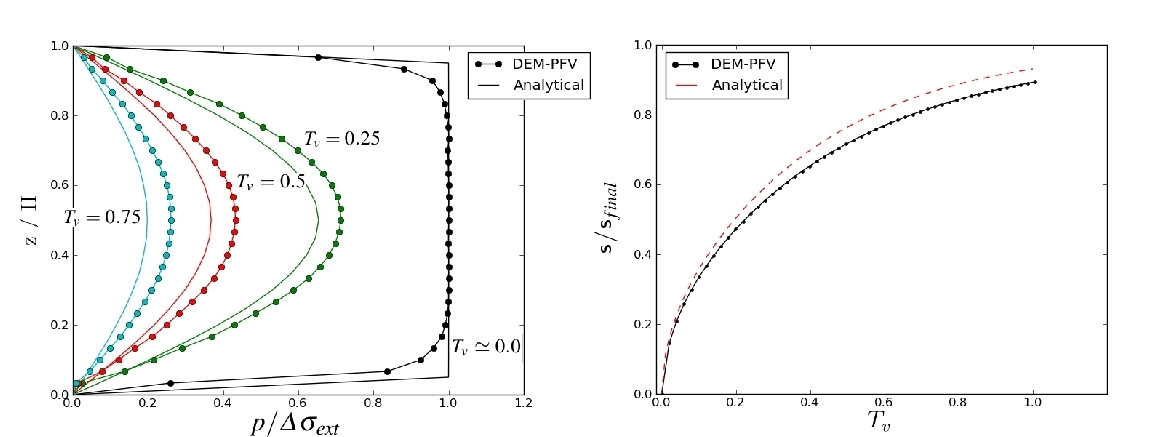}
\caption{Oedometric consolidation with large deformations -  the analytical solution is computed with the initial (small-strain) value $Cv$: evolution of pore pressure (left) and settlement (right).}
\label{Consol_HighCV1}
\end{figure}

\begin{figure}
 \centering
\includegraphics[width=140mm]{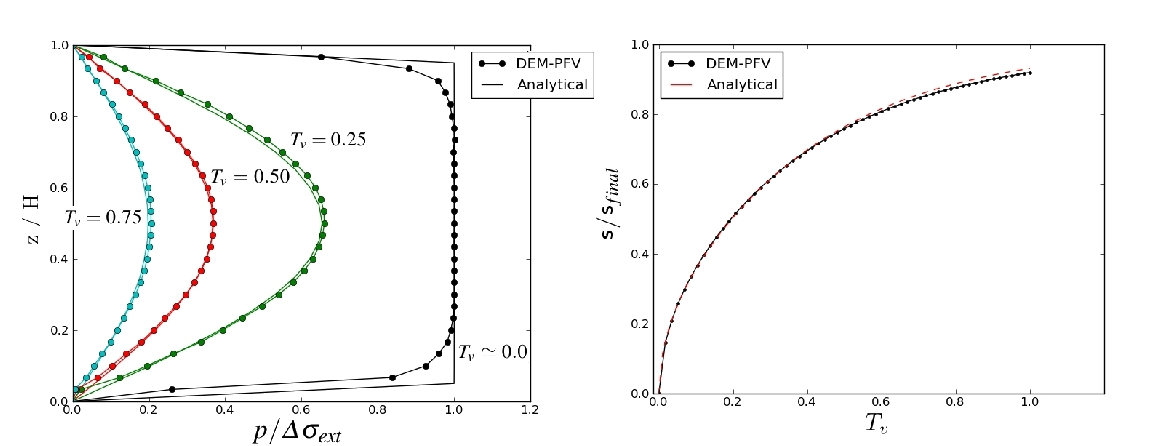}
\caption{Oedometric consolidation with large deformations - the analytical solution is computed with the final (secant) value of $Cv$: evolution of pore pressure (left) and settlement (right).}
\label{Consol_HighCV2}
\end{figure}

In Fig.\ref{Consol_HighCV1}, the analytical solution is computed by using the initial (tangent) value of $C_v$. Such choice leads to a discrepancy between the analytical and the numerical solutions. More precisely, the rate of deformation is underestimated. On the contrary, in fig.\ref{Consol_HighCV2}, the analytical solution is obtained with the secant value of $C_v$. The agreement is rather good in this case, although the difference remains larger than in the linear case (fig. \ref{Comparison_2}). This is because in case of large deformations the permeability is no longer constant, and the stress-strain behavior is no longer linear.
Well then, permeability changes and non-linearities discard the use of unique constants for the permeability and the oedometric modulus of the granular medium. The results obtained at large strain are thus picking out the limitation of the linear solution from Terzaghi.


\section{Immersed granular deposition}
\label{sec:deposition}
\begin{figure}
 \centering
\includegraphics[width=120mm]{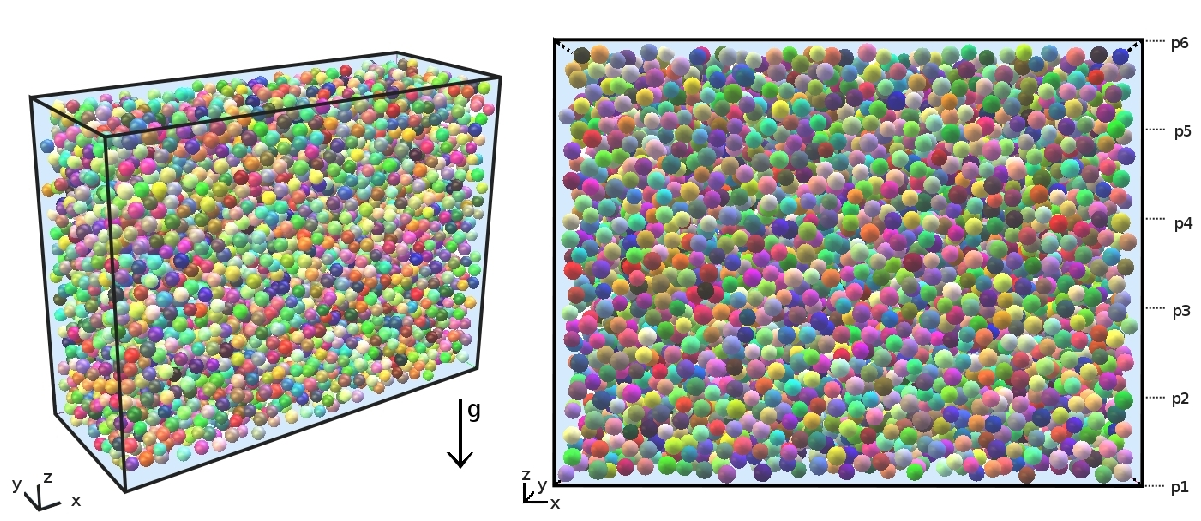}
\caption{Simulation of an immersed granular deposition (a). Position of fluid pressure sensors p$_i$ (b).}
\label{deposit}
\end{figure}

In this section the evolution of the fluid pressure and the effective stress is observed in a granular deposition problem. The simulation was set in order to simulate a fluid-filled vessel in which a number of spheres is placed and left to deposit under the action of gravity. No fluxes are allowed through lateral and bottom boundaries. A cloud of immersed spheres is created, as represented on fig.\ref{deposit}. Table \ref{tab:deposition} reports the parameters of the simulation. The fluid pressure is recorded at six different heights during the simulation, according to the scheme represented on fig.\ref{deposit}.

\begin{table}
\caption{Immersed granular deposition - Input data.}
\centering
\begin{tabular}{*{3}{c}}
\toprule
& INPUT DATA & \\
\midrule
Number of grains & [-] & 5000 \\ 
Sample dimensions & [m] & 2.00 x 1.50 x 0.75 \\
$\mu$ & [kPa $\cdot$ s] & 0.10 \\
$d_{50}$ & [m] & 0.06 \\
$\rho^s$ & [kg/m$^3$] & 2600 \\
$\rho^f$ & [kg/m$^3$] & 1000 \\
$p_{ext}$ & [kPa] & 0 \\
$E$ & [kPa] & 15000 \\
$k_t/k_n$ & [-] & 0.5 \\
\\
\bottomrule
\label{tab:deposition}
\end{tabular}
\end{table}

\subsection{Critical gradient}
Once the sedimentation process reaches a steady rate, the weight of the soil is completely carried by the fluid phase, that gets over-pressurized. These conditions are mechanically equivalent to a fluidization of the soil, which by definition consists in the cancellation of the effective stress as an effect of an upward ground water seepage. This situation occurs when the hydraulic gradient equalizes the so-called critical gradient
\begin{equation}
 \label{CritGrad}
i_c = \frac{\nabla p}{\gamma_f} = \frac{\gamma_{sat}-\gamma_f}{\gamma_f} 
\end{equation}
where $\gamma_{sat}$ indicates the specific weight of the mixture: $\gamma_{sat} = \gamma_{s}(1-n) + \gamma_f(n)$, with $\gamma_{s}$ and $\gamma_{f}$ the specific weight of the solid and the fluid phase, respectively. Considering the initial porosity of the suspension, $n = 0.61$, we can directly evaluate the critical gradient of the simulated suspension: $i_c = 0.624$.

\subsection{Simulation results}
\begin{figure}
 \centering
\includegraphics[width=140mm]{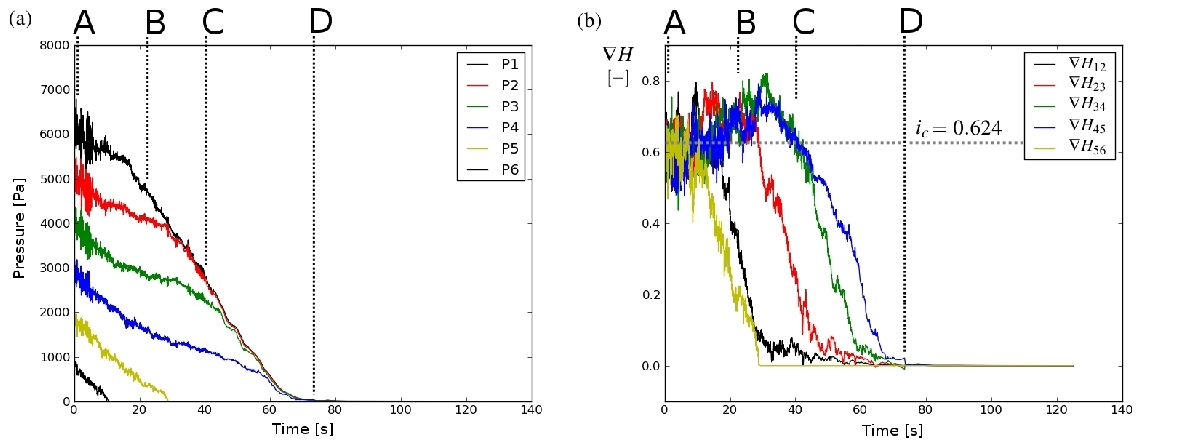}
\caption{Immersed granular deposition, case of tab.\ref{tab:deposition}. Fluid pressure measurements (a). Hydraulic gradient (b).}
\label{sensorDeposit}
\end{figure}

\begin{figure}
 \centering
\includegraphics[width=130mm]{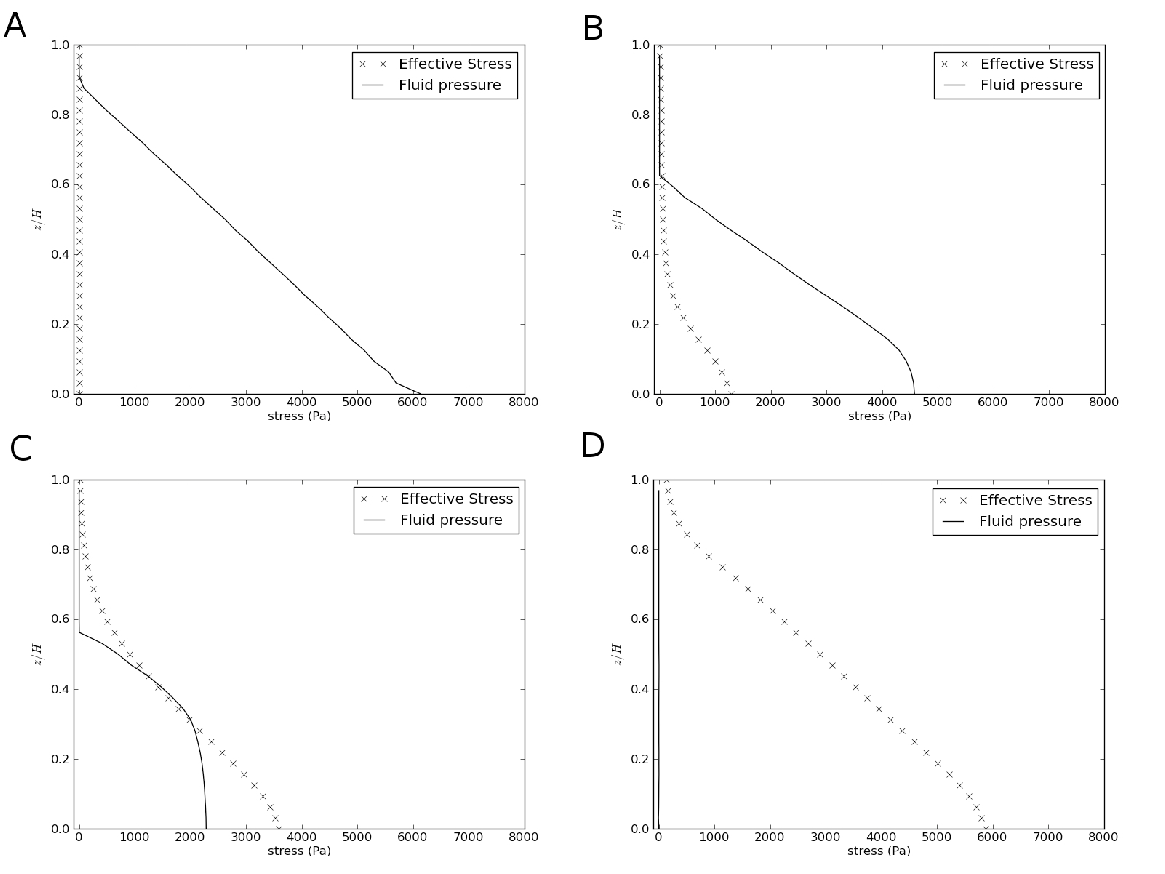}
\caption{Effective stress and fluid pressure evolution during the granular deposit.}
\label{StressDeposit}
\end{figure}

The result in terms of fluid pressure measured along the height of the vessel is plotted on fig.\ref{sensorDeposit}(a). The gradient of fluid pressure, evaluated for each couple of consecutive sensors, is plotted as well (b).
On fig.\ref{sensorDeposit}(a), when the pressure measured at a certain layer equals the one measured at an adjacent layer, it means that at that depth the spheres are newly in contact and the layer has stabilized. It can be observed, on fig.\ref{sensorDeposit}(b), how the initial values of the hydraulic gradient are close to the critical gradient defined previously, and finally goes to zero as the packing stabilizes. On fig.\ref{StressDeposit} a number of stress states, characterizing the evolution of the simulation, is represented. The initial load is entirely carried by the fluid phase, and the effective stress is none (A, see also fig.\ref{sensorDeposit} for the meaning of the capital letters). As soon as the particles start touching each other in the lower strata, the consolidation process starts, and the stress is transferred from the liquid phase to the solid skeleton (B-C). Finally, once the fluid pressure is fully dissipated and the consolidation process completed, the load is entirely carried by the solid skeleton (D).

These results highlight the flexibility of the coupling, which can handle the poroelastic couplings but also the transitions between a loose suspension and a solid state.

\subsubsection{Stokes and Reynolds numbers}
As introduced in section \ref{sec:fluid}, the soundness of the steady laminar flow hypothesis, in case of fluid-particles systems, is evaluated through the evaluation of the Stokes number. It is important to give an estimation of the Stokes number for the simulation of the granular deposition, especially for the initial configuration which is characterized by a relatively high value of porosity. From eq.\ref{stknumb2}, we may use the average distance between the particles (average throat diameter) as a characteristic length of the problem ($d$ in the equation). Noting as $e$ this distance and knowing the initial value of porosity $n=0.61$, we have:
\begin{equation}
 n = 1 - \frac{V_s}{V_t} = 1-\frac{4/3\,\pi d_{50}^3}{8(d_{50}+e)^3} = 0.61 \quad \rightarrow \quad e = 0.014m
\end{equation}
where $d_{50}$ is the mass-median-diameter of particles, $V_s$ and $V_t$ are the solid and the total volume, respectively.
Thus, we obtain:
\begin{equation}
 \label{stkImm}
S_t = \frac{\tau_c}{\tau_{\nu}} = \frac{\nu \tau_c}{e^2} = 500 >> 1
\end{equation}
where the order of magnitude of $\tau_c$ is estimated by looking at fig.\ref{sensorDeposit} how much time the lowest layers take to get stabilized, thus $\tau_c\simeq30s$. The Reynolds number can be also estimated by considering the permeability of the sample at the initial conditions, which is $K_d = 4.01621\cdot10^{-1}$ m/s, and the critical hydraulic gradient. We obtain $Re \simeq 10^{-2}$. The hypothesis of slow viscous flow holds for this simulation, thus validating the flow model assumptions.

\section{Conclusions}

The paper is devoted to the presentation of the DEM-PFV coupled model for viscous flow in granular materials. A key feature of the model is the description of the interaction between the solid and the fluid phases at the scale of pores and particles. This strategy, inspired by the pore-network approach, significantly reduces the computational cost generally associated to the direct simulation of flow in porous materials. Although the model is formulated in the framework of discrete mechanics, it results in a system of equations that is consistent with the theory of poroelasticity.

The ability of the model to solve transient problems in the quasi-static regime has been evaluated in the oedometer test simulation. The solution obtained, in the case of small deformations, is quantitatively in good agreement with Terzaghi's analytical solution, in terms of evolution of the excess pore pressure, stress and settlements, in time and space. In the case of large deformations, the limitations of Terzaghi's formulation due to the use of constant values of permeability and oedometric modulus have been highlighted. Micro-scale definitions of strain and stress have been introduced and allow meaningful interpretations of the results. Namely, the micro-scale stress can be considered as a micromechanical analogue of Terzaghi's effective stress (this may be also true to some extent in unsaturated materials, as suggested by \cite{Scholtes2009}).

To our knowledge, it is the first time that such quantitative agreement with the analytical solution of Terzaghi's problem is reported for a DEM-based hydromechanical model (although relatively good agreement was also found in \cite{Chen2007}). The reason may be that the implicit PFV formulation is based on strict incompressibility of the fluid, while the other coupling methods often use explicit formulations which need a finite (strictly positive) value of compressibility. The time-step of an explicit method depends on fluid compressibilty, so that capturing the poromechanical effects involved in the incompressible limit requires extremely small time-steps. Since, in addition, the accuracy of conventional methods (CFD,LBM,SPH) requires thousands of fluid DOFs per solid particle (while the number of PFV DOFs is of the order of the number of particles), the poromechanical effects may be recovered only at the price of inconveniently high computational costs.

Since the 1-D consolidation problem of Terzaghi is one of the simplest boundary value problems governed by poromechanical effects, we believe that it could be a standard problem for benchmarking coupling models. It could be argued that Terzaghi's problem is assuming incompressibility for the fluid while water, for instance, is not incompressible. However, the compressibility of water is so low that many experiments on soils are not influenced by it. Therefore, the numerical models should be able to produce results corresponding to the incompressible limit.

The numerical result that was obtained by simulating the granular deposition problem confirmed the robustness of the model and its ability to handle a wide range of solid-fluid interactions. Hence the strength of the DEM-PFV coupling is twofold. First, it is consistent with poromechanics although the granular material is defined only by its microscale geometry and contact laws. Second, the model is not restricted to a specific state of the materials, hence it naturally applies to study mechanisms that are out of the scope of conventional poromechanics, such as phase transitions \cite{Catalano2011a} or internal erosion \cite{Sari2011} in saturated granular materials. The current formulation is restricted to quasi-static regimes. The extension to inertial regimes is currently in development.

\section{Acknowledgements}
This study is part of the HydroFond project, with the support of the Department of Ecology, of the sustainable development, transports and housing, general direction of substructures, transports, and the sea, within the C2D2 program of RGCU. It was also supported by Grenoble Institute of Technology through the BQR-2008 program.

\bibliographystyle{abbrv}
\bibliography{biblio_full_condensed}

\end{document}